\documentstyle[12pt]{article}
 
\newlength{\extralineskip}
\newcommand{\beq}{\begin{equation}}
\newcommand{\eeq}{\end{equation}}
\newcommand{\bd}{\begin{displaymath}}
\newcommand{\ed}{\end{displaymath}}
\def\bea{\begin{eqnarray}}
\def\eea{\end{eqnarray}}
\def\nn{\nonumber}
\def\ba{\beq\new\begin{array}{c}}
\def\ea{\end{array}\eeq}

\def\inbar{\,\vrule height1.5ex width.4pt depth0pt}
\def\IC{\relax\hbox{$\inbar\kern-.3em{\rm C}$}}
\def\IR{\relax{\rm I\kern-.18em R}}
\def\IZ{{{\rm Z}\!\!{\rm Z}}}
\def\1{\relax{\rm 1\kern-.25em l}}
\def\pvint{{\int\!\!\!\!\!\!-}}

\def\Tr{{\rm Tr}}
\def\e{~{\rm e}}
\def\la{{\langle}}
\def\ra{{\rangle}}
\makeatletter
\newdimen\normalarrayskip              
\newdimen\minarrayskip                 
\normalarrayskip\baselineskip
\minarrayskip\jot
\newif\ifold             \oldtrue            \def\new{\oldfalse}
\begin{document}
\begin{titlepage}
\thispagestyle{empty}
 
\begin{flushright}
PUPT-1827  
\end{flushright}

\vskip1cm
\begin{center}
{\LARGE \bf Eigenvalue Dynamics and the Matrix Chain}\\
\vskip1cm
{\bf L.D. Paniak} \\
\bigskip
{\it Joseph Henry Laboratories, Department Of Physics \\
Princeton University \\
Princeton, New Jersey 08544, USA} \\
e-mail: paniak@feynman.princeton.edu
\vskip4cm

\begin{abstract}
We introduce a general method for transforming the equations of motion 
following from  a Das-Jevicki-Sakita Hamiltonian, 
with boundary conditions, into a boundary value
problem in one-dimensional quantum mechanics.
For the particular case of a one-dimensional  chain of interacting
$N\times N$ Hermitean matrices, the corresponding large $N$
boundary value problem is mapped into a linear Fredholm equation with
Hilbert-Schmidt type kernel. 
The equivalence of this kernel, in special cases, to a second order
differential operator allows us recover all previously known explicit
solutions for the matrix eigenvalues.
In the general case, the distribution of eigenvalues  
is formally derived through a series of saddle-point approximations.
The critical behaviour of the system, including a previously observed
Kosterlitz-Thouless transition, is interpreted in terms of the
stationary points.  In particular we show
that a previously conjectured infinite series of sub-leading critical
points are due to expansion about unstable stationary points 
and consequently not realized.
\end{abstract}

\end{center}
\end{titlepage}
\newpage
\setcounter{page}1
\section{Introduction}

The utility of studying the statistical mechanics of systems
which can be encoded in terms of matrix variables has
been evident over the past twenty years.  From the generalization
of Wigner and Dyson's work by Br\'{e}zin, Itzykson, Parisi and Zuber
\cite{bipz} to the use of matrix models in low dimensional gravity
(see \cite{ginsparg} for reviews), to modern implementations
of matrix strings \cite{mtrxstrings}, matrix models have proven
to be a convenient framework for calculation.  Remarkably,
throughout this period the dynamics of matrix models have been considered only 
in the `stationary' case where solutions of the equations of motion
are assumed to be time-independent. 
There arise circumstances though where the time evolution of 
a system governed by an action over matrix variables is relevant, especially
if there are definite initial and final conditions on the state of the
system. Examples include two-dimensional Yang-Mills
theory on the cylinder and sphere \cite{caselle,gromat}, matrix model
approaches to the continuous Hirota equation \cite{kazakov} and lattice
QCD \cite{kmm} and, the case we will focus on here, linear chains
of interacting Hermitean matrices.
It is our goal to give a coherent framework in which in to study this
class of matrix model problems.

When an action over matrix variables
can effectively be reduced to the eigenvalues, the 
treatment of the eigenvalues as a single, collective coordinate
\cite{sakita} provides a convenient viewpoint
for determining the Hamiltonian of the system.
For the matrix actions we will consider here the particular form is 
known as the Das-Jevicki-Sakita Hamiltonian \cite{jevicki}.  
The associated equations of motion 
are the non-linear Euler equations of fluid dynamics in 
one dimension and with the addition of appropriate
boundary conditions present a difficult mathematical problem.
Here we will show  how these boundary value problems can be 
interpreted in the familiar framework of the semi-classical limit of 
one-dimensional quantum mechanics. 
While in general solutions are still difficult to obtain,
we will focus our attention on the case  of 
a one-dimensional chain of interacting $N\times N$ Hermitean matrices
which was shown to be a Das-Jevicki-Sakita-type system by Matytsin \cite{mat1}
in the limit of large $N$.
In this situation a 
restatement of the equations of motion in terms of quantum mechanics
allows one to consider a linear problem which is most conveniently
stated as a Fredholm integral equation.  Formal solution of 
this equation in terms of iterations of the integral kernel, which can 
be evaluated by saddle-point methods, will
provide a framework for detailing the nature and position of critical 
points which arise in the system.

An interesting system from a statistical mechanical point of view,
the one-dimensional 
matrix chain was discussed previously as a model for string 
theory with a discrete target
space \cite{parisi}. In addition to this direct application,
the matrix chain is related by duality transformation to string theory on 
a circle with radius $R=1/a$ \cite{kleb1,kleb2}, string 
theory at a finite temperature \cite{boulatov}  and to   $O(2)$ 
sigma models coupled to two dimensional quantum gravity \cite{mat2,mat3}.
The model itself consists of Hermitean matrices $\phi_x$
on the sites $\{x \}$ of a one-dimensional lattice interacting with 
nearest neighbours.  The action of the system is
\beq
S_{V}=a \sum_{x} \Tr V(\phi_x) - \sum_{x}  \frac{ \Tr (\phi_x -\phi_{x+1})^2}
{2 a}
\label{mataction}
\eeq
where $a$ is the  lattice spacing and $V$ is the potential
at each site.  Assuming the   site potentials $V(\phi_x)$ to 
be analytic in $\phi_x$, the action can be reduced to the eigenvalues
of the matrix variables by  $SU(N)$ gauge rotations $\Lambda$
and the known result for the correlator \cite{iz}  
\beq
I[\phi_{x}, \phi_{x+1} ]=
\int d\Lambda \e^{-N \Tr \Lambda \phi_x \Lambda^\dagger \phi_{x+1}}
\eeq
Consequently, in the limit of large $N$,  the calculation of the partition
function reduces to determining solutions of equations of motion for 
the eigenvalues.
Despite its relative simplicity and importance in low-dimensional 
string theory and quantum gravity,
explicit solutions of the matrix chain
are only known for a few particular choices of potential $V$.  In Section 
Four we will show these cases arise when the equations of motion are 
equivalent to certain Sturm-Liouville eigenvalue problems.

This model exhibits a remarkable range of behaviour as one adjusts the 
lattice spacing, $a$.  For instance, in the limit of vanishing $a$ 
one recovers the   canonical example of a Das-Jevicki-Sakita system,
matrix quantum mechanics (restricted to a singlet sector)
\beq
S_{MQM}=  \int  dx~ \Tr [ \frac{1}{2} \dot{\phi}^2 (x)   +V(\phi(x)) ]  
\label{mtrxqm}
\eeq
In fact, the methods we will introduce in the next Section and use
to study the matrix chain find their most natural form when 
applied to such matrix mechanical systems on a 
finite interval $x \in [0,T]$.  

More interestingly, fixing the potential $V$ to satisfy $V^{\prime \prime}(0) =-4$,
at critical lattice spacing $a=1$, the matrix chain 
was argued to undergo a Kosterlitz-Thouless phase transition \cite{kleb1}.
The physics of this phase transition are most transparent in the 
dual picture of string theory on the circle.  Here strings wrapping around 
the non-contractible target space play the role of vortices which 
are liberated when the radius $R \sim 1/a$ becomes sufficiently small.
In terms of a discrete lattice, at sufficiently small lattice spacing $a<1$
the universal behaviour of the system is that of the continuous model
(\ref{mtrxqm}) due to communication between lattice sites.
When the spacing is larger than unity this communication is broken
and the system can be expected to factorize into a product of 
single $D=0$ matrix models, which in fact it does as $a
\rightarrow \infty$.  The main objective of studying the
matrix chain is to understand precisely how this change of 
character from $D=1$ matrix quantum mechanics to isolated $D=0$ matrix
models takes place at $a=1$.  

In addition to $a=0$ and $a=1$, it was argued in \cite{mat2} and
\cite{mat3} that at a discrete infinity of 
lattice spacings non-analytic behaviour arises in the distribution
of eigenvalues of the matrix variables $\phi_x$.  
Considering iterative approximation of a 
functional solution to the equations of motion,
it was argued these critical points arise
when $a =  \sin{ (p \pi /2 q)}$ where $p \leq q$ are each positive integers.
In Section Five, where we develop the general solution for the 
distribution of eigenvalues as a series of saddle-point integrations, 
we will show that this behaviour is not realized in the matrix chain.
In fact it will be shown that such critical behaviour arises
when unstable stationary points are mistakenly considered relevant 
in saddle-point approximations.  Moreover, in constructing the general
solution it will become clear that a countable infinity of unstable
stationary points appear in intermediate steps each of which can 
lead to misleading results.

In addition to the details just outlined, 
we recall the Kazakov-Migdal (lattice) model of induced 
QCD as an interesting application of the techniques developed here.
Since the $D$-dimensional model is effectively reducible to the one-dimensional
case, the non-trivial behaviour of the matrix chain demonstrated here
may help to overcome some of the apparent deficiencies of that
model of gauge interaction.

\section{Collective field theory to quantum mechanics}
\setcounter{equation}{0}

To begin we will demonstrate a non-linear change of variable
which allows the equations of motion derived from a  
Das-Jevicki-Sakita Hamiltonian \cite{jevicki}
\beq
{\cal H}[\rho,\Pi] = \frac{1}{2} \int dx~ \rho(x) \left\{ 
\left(\frac{\partial \Pi(x)}{\partial x} \right)^2 - \frac{\pi^2}{3} \rho^2(x)
\right\} + \int dx~ \rho(x) {\cal V}(x,t)
\label{djham}
\eeq 
to be determined as a temporal boundary problem in one-dimensional
quantum mechanics.
In (\ref{djham}),  $\rho(\lambda)$ is a probability distribution function which 
returns the fraction of eigenvalues of the matrix variables $\phi_x$ with 
value $\lambda$.
In the present context of collective field theory, $\rho$ 
plays the role of a coordinate and $\Pi$ is the  canonically 
conjugate momentum.
Here we have included a potential ${\cal V}$ which is most generally time-
dependent. Independent of time, (\ref{djham}) describes the  evolution
of the matrix quantum mechanics (\ref{mtrxqm}) with ${\cal V} =V$.   

The equations of motion following from this Hamiltonian are given 
by the functional variations with respect to the canonical variables
\beq
\frac{ \partial}{\partial t} \rho = \frac{ \delta {\cal H}}{\delta \Pi} 
~~,~~
\frac{ \partial}{\partial t} v = - \frac{\partial}{\partial x}
\left( \frac{ \delta {\cal H}}{\delta \rho} \right)
\eeq
where, for convenience, we have defined the `velocity' field
\beq
v(x,t) = \frac{\partial \Pi}{\partial x}
\eeq
Explicitly we see that the equations of motion are of the form 
of the Euler equations for a one-dimensional inviscid fluid 
experiencing a time-dependent force derived from ${\cal V}$
\bea
\frac{\partial \rho}{\partial t} +  \frac{\partial}{\partial x}
(\rho v) =0 \label{eqmo} \\
\frac{\partial v}{\partial t} + \frac{1}{2} \left(  \frac{\partial v^2}
{\partial x}
-\pi^2   \frac{\partial \rho^2}{\partial x} \right) = 
- \frac{\partial }{\partial x} {\cal V}(x,t)
\nonumber
\eea

As in all evolution problems we need to specify boundary conditions
to supplement the equations of motion.  In the case of matrix
quantum mechanics the classical treatment of \cite{bipz} assumed 
free boundary conditions for the evolution of the system over
infinite time.  In this case one is led to a study of the
ground state of the system and the equations of motion
(\ref{eqmo}) are solved for the stationary case of $v=0$.
In general though we are interested in arbitrary boundary conditions
for the evolution of the system through some time $T$.
To be definite we will specify the generic conditions on the 
fields $v$ and $\rho$ at time $t=0$ and $t=T$
\beq
\rho(x,0)= \rho_0(x) ~~~,~~~v(x,0) = v_0(x) 
\label{genbndconds}
\eeq
\bd
\rho(x,T)= \rho_T(x) ~~~,~~~v(x,T) = v_T(x)
\ed

We would now like to rewrite the equations of motion 
in terms of a single quasi-linear first order
differential equation.  Introducing the complex function
\beq
f(x,t) =   v(x,t) + i \pi \rho(x,t) 
\label{fdefn}
\eeq
it is easy to see that (\ref{eqmo}) leads to 
\beq
\frac{\partial f}{\partial t} + f \frac{\partial f}{\partial x} = 
- \frac{\partial }{\partial x} {\cal V}(x,t)
\label{burgers}
\eeq
For vanishing potential and real $f$
this is the Hopf equation which is the prototypical one dimensional
model of an equation which admits wave solutions complete with
shocks and other inherently non-linear behaviour \cite{whitham}.
We would like to solve this equation with the boundary conditions
given in (\ref{genbndconds}).  While the solution of the (complex) Hopf
equation is known explicitly \cite{mat1} in terms of initial data, typically
we are interested in cases where boundary data is supplied.
In these situations 
the explicit solution leads to an ill-posed, inverse-type problem
for the boundary conditions.  
Consequently we would like to find a more natural presentation of the problem. 
To this end we will 
consider a generalized  version of (\ref{burgers}) and introduce 
a dispersive term \cite{whitham} which will serve to smooth out the solutions and 
allow one to calculate the details of the solution $f$ in terms of boundary
conditions.  Adding such a term leads to a complex version of  the 
forced Burgers' equation 
\beq
i \hbar \frac{\partial^2 f}{\partial x^2} 
=\frac{\partial f}{\partial t} + f \frac{\partial f}{\partial x}
+ \frac{\partial }{\partial x} {\cal V}(x,t)
\label{NS}
\eeq
Obviously, in the limit as $\hbar \rightarrow 0$   the 
forced Hopf equation (\ref{burgers}) is recovered.  
The usefulness of this addition to the 
problem is that there exists a well-known
(to mathematicians) change of variable which will reduce 
(\ref{NS}) to a second order, linear differential equation.  This 
transformation is known as the Cole-Hopf transform and is given 
\beq
f(x,t) = - 2 i \hbar  \frac{\partial }{\partial x} \log{\psi(x,t)}
\label{psidefn}
\eeq
As is readily verified, this transformation leads (up to an irrelevant
additive constant of integration) to the Schr\"odinger equation
\beq 
H \psi =
\left[ - \hbar^2 \frac{\partial^2 }{\partial x^2} + 
\frac{{\cal V}(x,t)}{2} \right]
\psi = i \hbar \frac{\partial }{\partial t} \psi
\label{schroeq}
\eeq
Consequently we see that the time evolution of
the collective field Hamiltonian (\ref{djham}) is given by the semi-classical
$(\hbar \rightarrow 0)$
limit of one-dimensional quantum mechanics with the same potential, up to a 
factor.

Restricting to time independent potentials\footnote{This restriction is not essential.  In the general case the 
time evolution of the system  is given by a Dyson series expansion.}
${\cal V}(x,t) = {\cal V}(x)$,
the boundary conditions (\ref{genbndconds}) enter into the picture
as the start and end points of evolution by $H$ through time $T$
\beq
\psi(x,T) = \e^{-i H T/\hbar} \psi(x,0)
\label{operevol}
\eeq
where the wavefunction is defined in terms of the eigenvalue
density and velocity 
\beq
\psi(x,t) = \exp{\left[ - \frac{\pi}{2 \hbar} \int^x d \eta ~ \rho(\eta,t) + 
\frac{i}{2 \hbar} \int^x d \eta ~ v(\eta,t) \right] }
\eeq

Given the Green's function $G(x,u|T)$
for the Schr\"odinger problem it may be 
convenient to cast (\ref{operevol}) as an integral equation
\bea
\exp{\left[ - \frac{\pi}{2 \hbar} \int^x d \eta ~ \rho_T(\eta) + 
\frac{i}{2 \hbar} \int^x d \eta ~ v_T(\eta) \right] }& =&  
\label{geninteq} \\
\int du ~G(x,u|T) &&\!\!\!\!\!\!\!\!\!\!\!
\exp{\left[ - \frac{\pi}{2 \hbar} \int^u d \eta  ~\rho_0(\eta) + 
\frac{i}{2 \hbar} \int^u d \eta  ~v_0(\eta) \right] } \nonumber
\eea
In applications  the implementation of boundary conditions 
and Green's function varies depending on the situation.
For example, in investigations of the  
continuous Hirota equation \cite{kazakov} one is interested 
in the case of time-dependent potentials ${\cal V}(x,t)$ leading
to non-trivial Green's functions.  Here the initial and 
final velocities are specified  and the initial and final eigenvalue
distributions are identified ($\rho_0=\rho_T$) so that (\ref{geninteq}) takes
on the form of a non-linear integral equation for $\rho_0$. 

Another, more familiar application
is that of two-dimensional Yang-Mills theory \cite{caselle,gromat,thesis}.  
In this case the coordinates $(x,u)$ in 
(\ref{geninteq}) are replaced by their periodic counter-parts 
since the matrix variables are unitary rather than Hermitean.
The Green's function is that of the heat equation on a circle and 
the boundary conditions on $\rho$ and $v$
specify the topology of the two-dimensional space-time.
In general these cases again lead to  non-linear 
integral equations for the unknown fields but as we will now see,
there are situations where the general integral equation (\ref{geninteq})
leads to a tractable, linear equation.

\section{An application: the matrix chain}
\setcounter{equation}{0}

An application of the preceding formalism, which we will be investigating 
for the remainder of this Paper is that of the one-dimensional 
matrix chain. Instead of (\ref{mataction}), 
it will be convenient  to consider a different form of the action 
\beq
S_U = \sum_{x}  \Tr U(\phi_x) - \sum_{x}   \Tr \phi_x \phi_{x+1}
\label{mcaction}
\eeq
where we have rescaled the matrix field $\phi_x$ and the new site
potential $U$ is related to $V$ by
\beq
U(\phi_x) = a V( \sqrt{a} \phi_x) + \phi_x^2
\eeq
The evolution of eigenvalues in this system, in the limit of large $N$,
was shown by Matytsin \cite{mat1}
to be given by the Das-Jevicki-Sakita Hamiltonian (\ref{djham})
with vanishing potential 
${\cal V}$.  
Here the boundary conditions (\ref{genbndconds}), which play a crucial role, 
are given by
\bea
\rho_0(x) & \equiv & \rho(x,0) =  \rho(x,1) \label{bndconds}\\
v(x,0) &=& - v(x,1) = \frac{1}{2 } U^{\prime}(x) - x  \nonumber
\eea
The objective is to solve for the initial and final eigenvalue
density $\rho_0$ in terms of the potential $U$.

With ${\cal V}=0$, and  $\epsilon = i \hbar$,
the Schr\"odinger   equation (\ref{schroeq}) reduces to 
the linear heat equation.
Substituting the well known 
Green's function in (\ref{geninteq}) and 
implementing the boundary conditions (\ref{bndconds}), we find a linear
integral equation of second Fredholm type 
for the initial eigenvalue
density $\rho_0(x)=\rho(x,0)$ in terms of the effective potential
$U(x)$
\beq
\chi(x) = \frac{\lambda }{\sqrt{ 4 \pi \epsilon }} \int 
du \e^{\frac{- (x -u )^2}{4 \epsilon } + 
\frac{ x^2 - U(x) + u^2 -U(u)}{4 \epsilon}}  \chi(u)
\label{inteq}
\eeq
where
\beq 
\chi(x) = \exp \left[ -\frac{i \pi}{2 \epsilon}\int^x d \eta~ 
\rho_0(\eta) \right]
\label{phidefn}
\eeq
and we have introduced the constant $\lambda$ which does not affect
the asymptotic solution but is convenient for later discussion.
In this form (\ref{inteq}) has the natural interpretation
of an eigen-problem with eigenvector $\chi$   
and associated eigenvalue $\lambda$.
It is interesting to note that the same integral equation arose in 
the context of matrix chains previously \cite{parisi,kleb1}.
In these instances the authors were seeking the lowest
lying eigenvalues $\lambda$ in order to calculate the partition
function.  Here we are most concerned with the asymptotic 
$(\epsilon \rightarrow 0)$ form 
of the eigenfunctions $\chi$ in order to recover $\rho_0$.

Hence we have re-stated the differential form of the boundary value problem 
(eqns. (\ref{eqmo}) and (\ref{genbndconds}))
for the matrix chain as a well-posed,
linear problem for the eigenvalue density $\rho_0$. 
Moreover, the integral form (\ref{inteq}) allows one to write
equations for the eigenvalue density immediately. Formally,
in the limit of vanishing $\epsilon$, the integral in (\ref{inteq})
can be calculated
by saddle-point methods. This calculation is particularly 
straightforward
when only a single stationary point, $u_s$ contributes to 
the integral (\ref{inteq}). This stationary point is determined by the 
extrema condition  
\beq
x= U^{\prime}(u_s)/2 +   i \pi \rho_0(u_s) 
\label{first} 
\eeq
Likewise, at this stationary point $\rho_0$ can be formally evaluated using its
relation to $\chi$ from equation (\ref{phidefn}) 
\beq
u_s= U^{\prime}(x)/2 -  i \pi \rho_0(x)  
\label{second}
\eeq 
Labeling $x$ by $G_+$ in equation (\ref{first}) and $u_s$ by 
$G_-$ in equation (\ref{second}), we see the saddle-point 
evaluation of the integral equation (\ref{inteq}) leads to 
Matytsin's solution \cite{mat1} of the boundary value problem
\beq
x= G_+( G_-(x)) =  G_-( G_+(x))
\label{matrelns}
\eeq
This form of the solution of the problem is well-suited to detailed
local analysis as demonstrated in \cite{mat2,mat3}.
However, for a global analysis which can accommodate the existence of 
multiple stationary points, we will 
take the (well-studied) integral form (\ref{inteq}) over the 
functional equation (\ref{matrelns}).  
Before considering the particular characteristics of solutions though,
we will utilize the interpretation of integral operators as self-adjoint
operators to recover all
previously known explicit solutions $\rho_0$ of the matrix chain.

\section{Integral and differential operators}
\setcounter{equation}{0}

An analysis of an integral equation centers on the properties of the 
kernel $K$.  In our case, the kernel of (\ref{inteq}) is 
\beq
K(x,u) = \frac{1}{\sqrt{ 4 \pi \epsilon}} \e^{  \frac{ xu}{2 \epsilon} - 
\frac{ U(x) + U(u) }{4 \epsilon} }
\label{kdefn}
\eeq
This positive definite kernel is real, symmetric  and hence
Hermitean.  
In the following we will assume the kernel to have many convenient
properties which follow from the boundedness of all $n$-fold iterations
\beq
\int dx_1 \cdots dx_{n-1} K(x_1,x_2) \cdots K(x_{n-1},x_n) < \infty
\eeq
Consequently we will only consider potentials $U(x)$ which 
increase at infinity faster than $x^2$.
Under such conditions $K$ defines a compact operator and
represents a self-adjoint operator 
on the space of square-integrable functions. From classical operator theory
\cite{inteqrefs}
there is a countable infinity of orthogonal eigenfunctions 
$\{ \chi_n(x) \}$ which form a complete 
set on this space.   It is our goal to obtain, as a function of the site potential $U$, the asymptotic form of these eigenfunctions which determine
the large $N$ solution of the eigenvalue distribution $\rho_0$
through (\ref{phidefn}).

There are a number of cases in which the kernel $K$ is equivalent to 
familiar self-adjoint operators
in physics. In fact only in these instances are explicit
solutions of the large $N$ matrix chain known.
In each case the eigenvalue 
distribution $\rho_0$ is the solution of a quadratic equation.  
Here we
will show that such situations arise by considering the $\epsilon \rightarrow
0$ limit of second order self-adjoint differential operators 
${\cal L}$ for which ${\cal L}$ and $K$ share a complete set of eigenfunctions 
and define the same action on the Hilbert space. 

In order to determine the conditions that  ${\cal L}$ must
satisfy to be equivalent to the integral kernel we act
on $\chi_n$ as defined by the integral equation (\ref{inteq})
\beq
E_n \chi_n(x) = {\cal L}_x \chi_n(x) = 
\int du ~ {\cal L}_x K(x,u) \chi_n(u) 
\label{pf1}
\eeq
Likewise, acting under the integral sign and using the self-adjoint 
property of ${\cal L}$
\beq
E_n \chi_n(x) = \int du~ K(x,u) {\cal L}_u \chi_n(u) = 
\int du  ~\chi_n(u) {\cal L}_u K(x,u)
\label{pf2}
\eeq
Subtracting (\ref{pf2}) from (\ref{pf1}) we find that
${\cal L}$ will be equivalent to the integral
kernel $K$, if and only if we satisfy the natural commutativity condition 
\beq
( {\cal L}_x - {\cal L}_u ) K (x,u) = 0
\label{Lreq}
\eeq
It follows from the symmetry of the kernel that the action of ${\cal L}$
generates symmetric functions of $x$ and $u$ which will
cancel out in (\ref{Lreq}),
\beq
{\cal L}_x K(x,u)= h(x+u, xu) K(x,u) = \sum_{mn} a_{mn} (x+u)^m (x u)^n
K(x,u)
\label{symmcond}
\eeq
where $h$ is an arbitrary function of two variables with Taylor expansion
coefficients $\{a_{mn} \}$.  
 
This construction is valid for any differential operator but for simplicity 
we will consider only second order differential operators of 
Sturm-Liouville form
\beq
{\cal L}_x = 
-4 \epsilon^2 \frac{d}{dx} \left( a_2(x) \frac{d}{dx} \right) + a_0(x)
\label{sturmliou}
\eeq
Now it is a simple matter to find the potentials $U$ and associated coefficients $a_2$ and $a_0$ which
are consistent with the symmetry condition (\ref{symmcond}). 
Since a second order ${\cal L}$
can only produce quadratic polynomials in $u$, (\ref{symmcond}) leads to the equation
\beq
-a_2(x)( u - U^\prime(x)/2)^2 - 2 \epsilon
\left(a_2(x)(u-U^\prime(x)/2) \right)^\prime + a_0(x) 
=\sum_{m+n \leq 2} a_{mn} (x+u )^m (x u)^n 
\eeq
Expanding each side of the second equality
in powers of $u$ and equating coefficients gives, to leading order in $\epsilon$,
the result
\bea
U^\prime(x) &=& - \frac{ a_{11} x^2 + (2 a_{20} + a_{01} )x + a_{10} }
{a_{02} x^2 + a_{11} x + a_{20} } \label{gen2soln}\\
a_2(x) &=& -( a_{02} x^2 + a_{11} x + a_{20}) \nonumber \\
a_0(x) &=& a_{00} + a_{10} x + a_{20} x^2 - 
\frac{ (a_{11} x^2  +  (a_{01} + 2 a_{20}) x + a_{10})^2}
{4 ( a_{02} x^2  + a_{11} x + a_{20} )} \nonumber
\eea
Having determined the coefficients $a_2$ and $a_0$, we can solve 
for the explicit form of the associated eigenvalue distribution.
The differential form of the integral equation (\ref{inteq}) is
\beq
{\cal L}_x \chi(x) = E \chi(x)
\label{diffform}
\eeq
where the eigenvalue $E$ is related to that of the integral form 
by $\lambda= \e^{E/\epsilon}$.
In the limit of vanishing $\epsilon$, the solution of this differential
equation is given simply by the WKB approximation which implies 
the replacement of derivatives with `momenta'
\beq
i \epsilon \frac{d \chi}{dx} \rightarrow \frac{ \pi}{2 } \rho_0
\eeq
With this substitution, (\ref{diffform}) reduces to a quadratic
equation in the eigenvalue distribution which is solved by
\bea
\rho_0(x) &=& \frac{1}{\pi}  \sqrt{\frac{E- a_0(x) }{a_2(x)} } 
\label{genrho}\\
&=&  \frac{ 
\sqrt{4 ( a_{02} x^2  + a_{11} x + a_{20} )( a_{00} + 
a_{10} x + a_{20} x^2 -E) -(a_{11} x^2  +  
(a_{01} + 2 a_{20}) x + a_{10})^2 }}
{2 \pi ( a_{02} x^2  + a_{11} x + a_{20} )} \nonumber
\eea
Since $\rho_0$ is to be interpreted as a probability distribution 
describing the eigenvalues of the matrix variables, it must be properly
normalized. This is carried out by integrating $\rho_0$ over the
positive support of $(a_0 -E)/a_2$ and fixing the undetermined 
constant $E$ to satisfy 
\beq
\int dx \rho_0(x)  =1
\eeq

Integrating (\ref{gen2soln}), 
we find the most general potential of the matrix chain which 
is consistent with a second order, self-adjoint differential operator
\bea
U(x)&=& - \frac{a_{11}}{ a_{02}} x - 
\frac{2 a_{02}^2 a_{10} - a_{01} a_{02} a_{11} + a_{11}^3 - 2 a_{02} a_{11} a_{20}}
{a_{02}^2 \sqrt{ 4 a_{02} a_{20} - a_{11}^2} } 
\arctan{\left[ \frac{a_{11} + 2 a_{02} x}
{\sqrt{ 4 a_{02} a_{20} - a_{11}^2} } \right]} \nn \\
&&
- \frac{a_{01} a_{02} - a_{11}^2}{2 a_{02}^2} 
\log{\left[ a_{20} + a_{11} x + a_{02} x^2 \right]}
\label{gen2pot}
\eea
Contained in this general form are a number of previously examined 
examples of the matrix chain.
First, setting $a_{11}=0$ and $a_{10}=0$
recovers the symmetric double Penner-type potential first considered
by Matytsin \cite{mat1}. Also the asymmetric
double Penner
considered in connection with the Kazakov-Migdal model of induced QCD 
\cite{penner} can be recovered (see \cite{chekhov} for recent work).
In each of these cases the eigenfunctions $\chi$ are related to 
prolate spheroidal wave functions.
Additionally, with $a_{20}=1$, 
the limit $a_{02} \rightarrow 0$ recovers the quadratic potential case solved 
by Gross \cite{gross} and Makeenko \cite{makeenko}. 
With $U^\prime = (a_{01} +2) x \equiv  2 m^2 x$,
${\cal L}$ is the Schr\"{o}dinger operator for the simple harmonic
oscillator and the general expression for the
distribution of eigenvalues (\ref{genrho}), properly normalized, reduces to
\beq
\rho_0(x) = \frac{1}{\pi} \sqrt{ 2 \sqrt{m^4 -1}- (m^4 -1) x^2 }
\label{gaussans}
\eeq
 
In each of these cases the analytic structure of the
eigenvalue distribution (\ref{genrho}) is what one would 
expect of a $D=0$ one-matrix model with a particular potential. 
This correspondence was previously suggested for arbitrary potential in the 
matrix chain, but such a simple solution of the integral equation (\ref{inteq})
is not possible.
In fact with further computation it can be argued that higher, finite, order 
self-adjoint differential operators are equivalent to the integral kernel 
$K$ only if they are functions of the second order operator ${\cal L}_x$
we have constructed above.  
Consequently, the large $N$ eigenvalue distribution $\rho_0$ of the 
matrix chain is the solution of a polynomial equation only for potentials
of the form (\ref{gen2pot}). This suggests that for a generic potential
the solution of the matrix chain problem follows from the 
semi-classical approximation to a pseudo-differential operator
as suggested by naively extending the results for 
finite chains of matrices \cite{douglas}. We will not pursue this line 
of reasoning here but return in the next Section to the integral 
form (\ref{inteq}) and give a method
for obtaining the general solution and describe its unique features.
 
\section{General solution and classification of critical points}
\setcounter{equation}{0}
 
A phase transition in a system is characterized 
by a change in the analytic structure of observables as functions of 
external parameters.
In standard matrix models this is commonly taken to include 
a change in analytic structure of the 
distribution of eigenvalues\footnote{This assumption of a physical phase 
transition following from a change in the analytic structure of the eigenvalue 
distribution does not always hold \cite{po}.  Nevertheless, in the absence of 
calculating physical observables, we will take a change in structure of the 
eigenvalue distribution to be a strong hint of a physical transition.}.  
In the present case we have shown that the 
distribution of eigenvalues for a linear chain of Hermitean matrices is
determined completely by a linear integral equation (\ref{inteq})
involving the kernel
$K$ (\ref{kdefn}). In this Section we will go further and 
solve for the asymptotic eigenfunctions $\chi$
as a series of iterations of $K$. Given that iterations of $K$ 
are calculable by saddle-point methods and the relation
(\ref{phidefn}), we can effectively solve for the 
distribution of eigenvalues $\rho_0$.  In particular we will
be able to determine the critical structure of the matrix
chain by analyzing changes in the analytic structure of the 
kernel $K$ and its iterations.

We begin by slightly generalizing   the problem of finding the eigenfunctions
$\chi$ in order to make use  of well-known techniques \cite{inteqrefs}.
Instead of the integral equation (\ref{inteq}) let us consider for a moment
an inhomogeneous version with $c$ some real constant
\beq
\chi(x) = c + \lambda \int du ~K(x,u) \chi(u) \equiv c + \lambda K \chi
\label{nonhomg}
\eeq
Since there are no constant eigenfunctions of the kernel $K$,
the unique solution of (\ref{nonhomg}) is given by the Neumann series
which builds the solution up through an iterative procedure
\bea
\chi_0 &=& c \\
\chi_1 &=& c + \lambda K \chi_0 \nn \\
&\vdots& \nn \\
\chi_n &=& c + \lambda K \chi_{n-1} \nn
\eea
Taking the limit of this process generates the solution
\beq
\chi(x) = \left( 1 + \sum_{n=1}^\infty \lambda^n \int du ~ K^n(x,u) 
\right) c 
\equiv R(x;\lambda) c
\label{rdefn}
\eeq
where $R$ is commonly referred to as the resolvent kernel and
$K^n$ is given by the convolution of $n$ kernels
\beq
K^n(x,u) = \int dz_1 \cdots dz_{n-1} K(x,z_1)K(z_1, z_2) \cdots
K(z_{n-1},u) 
\eeq
The convergence of  the resolvent kernel is 
guaranteed by the boundedness of $K^n$ and the Hilbert-Schmidt theory of 
integral equations \cite{inteqrefs}.
In order to recover the homogeneous solution we should set $c=0$ which
leads to a trivial solution for $\chi$. Fortunately for us only the $c$ independent logarithmic
derivative of $\chi$ is required  to recover the eigenvalue distribution 
$\rho_0$ of the matrix chain.
From the definition of $\chi$, (\ref{phidefn}) and (\ref{rdefn})
we have the result
\beq
\rho_0(x) = \lim_{\epsilon \rightarrow 0} 
\frac{2 i \epsilon}{\pi} \frac{d}{dx} \log{ \chi(x)}
= \lim_{\epsilon \rightarrow 0} \frac{2 i \epsilon}{\pi} \frac{d}{dx} \log{R(x;\lambda)}
\eeq

Hence  we can extract the eigenvalue
distribution from the resolvent kernel $R$, which is expressible solely in 
terms of iterations of the kernel $K$ as defined in (\ref{rdefn}).
This is an important observation  since the iterated 
kernels are calculable in the limit of vanishing $\epsilon$.
Explicitly, the $n^{th}$ iterated kernel can be written for $n \geq 2$
\beq
K^n(x,u)= \frac{1} { (4 \pi \epsilon)^{n/2}}
\e^{- \frac{ U(x) + U(u)}{4 \epsilon}} 
\int dz_1 \cdots dz_{n-1} \e^{- S/2 \epsilon} 
\label{kndefn}
\eeq
where the `action' $S$ is given by
\beq
S = \sum_{i=1}^{n-1} U(z_i) - \sum_{i=1}^{n-2} z_i z_{i+1} - z_1 x - z_{n-1} u
\label{sdefn}
\eeq

Evaluating the integral over $\vec{z}=\{ z_1, \ldots, z_{n-1} \}$ is a 
straightforward application of 
saddle-point methods in the limit of vanishing $\epsilon$.
In this limit the dominant contributions to the integral come from 
the vicinity of the stationary points of the function $S$  
\beq
\left. \frac{\partial S}{\partial z_i} \right|_{\vec{z}^\alpha} = 0
\eeq
In general there are many such stationary points, which we will label
by $\alpha$.  Expanding $S$ up to quadratic order about the stationary
points, the resulting Gaussian integrations give the approximation
to the iterated kernel
\beq
K^{n+1}(x,u) = \frac{1} { \sqrt{4 \pi \epsilon} }
\e^{- \frac{ U(x) + U(u)}{4 \epsilon}} 
\sum_{\alpha} \frac{(-1)^{\lambda_n(\vec{z}^\alpha  )}}
{ \sqrt{\det{ {\cal H}_n(\vec{z}^\alpha )}} } 
\e^{- S(\vec{z}^\alpha )/2 \epsilon}
\label{saddlek}
\eeq
where $\lambda_n(\vec{z}^\alpha)$ is the number of negative eigenvalues of the 
$n \times n$ Hessian ${\cal H}_n$ of second derivatives of $S$
evaluated at the stationary point
\beq
{\cal H}_n (\vec{z}^\alpha )
= \left. \frac{ \partial^2 S}{ \partial z_i \partial z_j} 
\right|_{\vec{z}^\alpha}
= \left( \begin{array}{ccccc} U^{\prime \prime} 
(z_1^\alpha) & -1 & 0 & 0 & \cdots \\
-1 & U^{\prime \prime} (z_2^\alpha) & -1 & 0& \cdots \\
 & \cdots & \cdots & \cdots 
 \\
0 & \cdots & 0 &-1 & U^{\prime \prime} (z_{n}^\alpha) 
\end{array} \right) 
\eeq
While we have counted contributions from all stationary points of $S$
in (\ref{saddlek}), the only relevant ones are those 
with minimum `action'.
The contributions of other, irrelevant, stationary points 
are exponentially suppressed in the limit of vanishing $\epsilon$.
 
Unfortunately it is difficult to calculate the iterated kernels 
and the resolvent in closed form to obtain explicit results
for the eigenvalue distribution.  It is possible though to 
extract enough qualitative information from (\ref{saddlek})
to outline a one-dimensional phase diagram of the matrix chain.
To be definite we will consider a quartic  potential
\beq 
U(\phi) = \phi^4 + m^2 \phi^2
\eeq
This simple form is convenient since it allows a clear view of 
the intrinsic behaviour of the matrix chain which depends on 
the strength of the quadratic term.

Starting with $m^2$ large and positive we expect that the eigenvalue
distribution 
will be localized near the origin where the quadratic term of the potential
is dominant.
Consequently, the solution (\ref{gaussans}) for the pure quadratic 
potential is a good approximation to the true solution and is 
smooth deformation of it.  
Making contact with the saddle-point evaluation of the iterated kernels
(\ref{saddlek})
it is easy to convince oneself that for large $m^2$ there is only 
one stationary point of the action (\ref{sdefn}) and it is located near the
origin $\vec{z}=0$.
As $m^2$ is decreased, the solution (\ref{gaussans})
of the pure quadratic potential becomes unstable at $m^2 =1$ and 
in fact this instability is present in the full solution.
Information about the stability of stationary points 
is contained in the eigenvalues $\mu_r$ of the 
Hessian ${\cal H}_s(z)$, which evaluated at the 
origin in $z$-space give
\beq
\mu_r = 2 m^2 - 2 \cos{ \frac{r \pi}{s +1} } ~~~,~~~ r=1,\ldots,s
\label{hessevals}
\eeq
Consequently we see that when evaluating arbitrarily high iterations
of the kernel, $K^n$, the Hessian at the origin
will being to develop negative eigenvalues at $m^2=1$ and this stationary
point becomes unstable.  Since $S$ is bounded from below for our choice of 
potential,  there are other stationary points
where $S$ has a lower magnitude and it is these which will 
dominate the integral (\ref{kndefn}) for $K^n$.

This degeneration at $m^2=1$ signals a phase transition in the 
matrix chain and the physical meaning of it can be found 
by returning to the original form of the matrix chain 
action (\ref{mataction}).  The potential there, $V$ is 
related to the current $U$ by the one dimensional lattice
spacing $a$ through
\beq
U(\phi) = a V( \sqrt{a} \phi) + \phi^2
\eeq
Taking two derivatives with respect to $\phi$
and fixing (without loss of generality)
$V^{\prime \prime}(0)=-4$ to agree with the conventions of  \cite{mat2}, we find
\beq
m^2= 1 -2 a^2
\label{utov}
\eeq
Hence, the instability at $m^2=1$ corresponds to the limit of vanishing
lattice spacing and we are observing a phase transition to $D=1$ matrix
quantum mechanics (i.e. (\ref{mtrxqm})) from the $m^2>1$ phase where 
the solutions are qualitatively what one would expect from a $D=0$
single matrix model.
In fact this observation can 
be made more concrete by returning to the integral equation (\ref{inteq}) 
but in terms
of the potential $V$ and lattice spacing $a$,
\beq
\e^{ -\frac{i \pi}{2 \epsilon}\int^x d \eta~ 
\rho_0(\eta) }
 = \frac{\lambda }{\sqrt{ 4 \pi \epsilon }} \int 
du \e^{\frac{- (x -u )^2}{4 a \epsilon } - a
\frac{   V(x) + V(u)}{4 \epsilon}}  \e^{ -\frac{i \pi}{2 \epsilon}
\int^u d \eta~ \rho_0(\eta) }
\label{d1inteq}
\eeq
It is a straightforward calculation to show that for vanishing $\epsilon$,
and to leading order in vanishing $a$, (\ref{d1inteq}) is
solved by 
\beq
\rho_0(x) = \frac{1}{\pi} \sqrt{ E - V(x)}
\eeq
which is the well-known solution for the ground state of matrix quantum
mechanics \cite{bipz}.

As $m^2$ is decreased further below this transition point, the stationary
point of $S$ at the origin $\vec{z}=0$
is seen to become progressively more unstable as the number of 
negative eigenvalues of the Hessian (\ref{hessevals}) grows.
Here the stationary points relevant to the saddle-point integrations
(\ref{saddlek}) are not in the neighbourhood of 
the origin and are stable under changes in $m^2$.  The only instability
that occurs with the relevant stationary points in this phase is due to a
$\IZ_2$ symmetry of the action $S$.  Recalling its definition
\beq
S = \sum_{i=1}^{n-1} U(z_i) - \sum_{i=1}^{n-2} z_i z_{i+1} - z_1 x - z_{n-1} u
\eeq
we see that for vanishing $x$ and $u$ there is a symmetry under $z_i \rightarrow
-z_i$.  It follows that a relevant stationary point at $\vec{z}^*$ has a
mirror image at $-\vec{z}^*$ and as $x$ and $u$ are tuned through the 
origin the relevant stationary point can change from one to the 
other.  In this way non-analytic behaviour arises in the solution of 
the eigenvalue distribution $\rho_0$, presumably of the 
form 
\beq
\rho_0 \sim |x|
\eeq
which is consistent with the critical behaviour of $D=1$ matrix quantum
mechanics.  Of course this detail is difficult to verify without 
explicit calculation.

In addition to this standard behaviour it was argued in \cite{mat2}
that there exist a countable infinity of additional points in this
phase where the eigenvalue distribution $\rho_0$ exhibits 
sub-leading non-analyticities.  In fact we can reproduce these results
if we assume, incorrectly, that the origin $\vec{z}=0$
is a relevant stationary point
in the evaluation of $\rho_0$.
Considering the vanishing  of Hessian eigenvalues (\ref{hessevals}), 
we see that the naive saddle-point approximation to $K^{s+1}$ (\ref{saddlek})
will break down when 
\beq
m^2 = \cos{ \frac{r \pi}{s +1} }
\label{vanevals}
\eeq
It is well known that at degenerate stationary points there is a 
change in the analytic structure of a saddle-point approximated 
integral so it is natural to assume that when (\ref{vanevals}) is 
satisfied, $K^{s+1}$ and the eigenvalue distribution $\rho_0$ will
exhibit non-analyticities.
In terms of lattice spacing the degeneracy condition (\ref{vanevals})
can be found by comparing with (\ref{utov}).  The result
\beq
a= \sqrt{ \frac{1- \cos{ \frac{r \pi}{s +1}}}{2} } = \sin{\frac{r \pi}{2 (s +1)}}
\label{vana}
\eeq
with $r$ and $s$ ranging over positive integers is exactly the condition
for subleading critical behaviour as found in \cite{mat2}.
Again we stress that this behaviour is not realized in the matrix chain
and only arises when one takes into account unstable stationary points
in the evaluation of saddle-point integrations.  This is the main
shortcoming of the functional form (\ref{matrelns}): 
in general there are a countable 
infinity of solutions and it is a difficult task to discover 
the true solution with minimum action.

There is one more phase transition in the matrix chain as $m^2$ decreases
further.  From (\ref{hessevals}) we see that as $m^2 \rightarrow -1$ 
all eigenvalues
of the Hessian at the origin become negative and this stationary point
is degenerate in all directions.  On the surface, besides this observation
of an irrelevant stationary point, there is little change in the system.
The relevant stationary points are found by solving polynomial equations
and their stability is determined locally as a smooth function of $m^2$.  
This is misleading though since $m^2=-1$ corresponds 
to a lattice spacing of unity which, in our scaling, is where a
Kosterlitz-Thouless phase transition was shown to occur \cite{kleb1,boulatov,
mat2}.  The only overt sign of the transition here
is that the integral kernel $K$ again changes analytic structure to
\beq
K(x,u)= \frac{1}{\sqrt{4 \pi \epsilon}} \e^{\frac{ (x+u)^2}{4 \epsilon} -
\frac{x^4 + u^4}{4 \epsilon} }
\eeq
Near the origin it has the form of a translation kernel which is of 
a different character than the original kernel (\ref{kdefn}).  
If we consider the higher, quartic
terms to only make the kernel well-defined, the behaviour of the 
eigenvalue distribution $\rho_0$ can be found by restricting 
the range of integration in the integral equation (\ref{inteq})
\beq
\e^{- \frac{i \pi}{2 \epsilon} \int^x d \eta ~ \rho_{KT}(\eta) } 
= \frac{1}{\sqrt{4 \pi \epsilon}} \int_{-b}^b  du~
\e^{\frac{ (x+u)^2}{4 \epsilon} } 
\e^{- \frac{i \pi}{2 \epsilon} \int^u d \eta ~ \rho_{KT}(\eta) }
\eeq
In this form it is clear that the solution will depend only on the 
constant $b$ which sets a scale \cite{mat2}, otherwise it is 
a universal quantity.  Unfortunately the asymptotic $(\epsilon \rightarrow 0)$ 
solutions of this equation are not known and so remains the 
eigenvalue distribution $\rho_{KT}$
of the matrix chain at the Kosterlitz-Thouless transition.

Finally, for $m^2<-1$ there persists a pair of $\IZ_2$ related 
stationary points that at $u$ and $x$ vanishing trade off
relevance.  It is conjectured that this non-analytic behaviour
is what one would find in a $D=0$ single matrix model, i.e.
\beq
\rho_0 \sim |x|^2
\eeq
It would seem that the same mechanism which produced the critical
exponents of matrix quantum mechanics for $-1<m^2<1$ produces
different exponents here.  This may be because of the total degeneracy
of the origin but explicit calculations of the iterated kernels
(\ref{saddlek}) and the eigenvalue distribution $\rho_0$ are
needed to make definite statements about the characteristics
of the solutions of this phase of the model.

\section{Conclusions and an application}
\setcounter{equation}{0}
 
To review, we have demonstrated a general technique for 
expressing the equations of motion for systems whose time
evolution is governed by Das-Jevicki-Sakita Hamiltonians with boundary
conditions as an evolution problem in one-dimensional quantum
mechanics.  Specifying to the particular example of the 
one-dimensional chain of interacting Hermitean matrices we
showed that in this case the general formalism reduces
to a linear integral equation for the large-$N$ eigenvalue
density of the matrices.  Analyzing the associated integral 
kernel we were able to recover all previous explicit solutions
of the matrix chain from asymptotic solution of particular
second order ordinary differential equations.  In the general
case, the solution of the integral equation was developed in 
terms of iterations of the kernel and from this solution 
the critical behavior of the system was found and 
a universal integral equation for the distribution of 
eigenvalues at the Kosterlitz-Thouless point was presented.
In addition it was demonstrated that the sub-leading 
critical behaviour observed in \cite{mat2} results from 
contributions from unstable stationary points in saddle-point
approximations and hence is not observed.

What was not accomplished here is the explicit calculation of 
a non-trivial solution to the large-$N$ matrix chain using 
the formalism developed here.  It would be nice to find a potential
for which the series of iterated kernels could be explicitly calculated
and summed and the eigenvalue distribution extracted.  This is of
particular interest in testing the conjectures involving the 
existence of a Kosterlitz-Thouless phase transition in the system.
With a global solution for $\rho_{KT}$ one would be in a position
to calculate explicitly the free energy of the matrix chain 
and check the critical behaviour. Of course only the leading 
non-analytic part
of the free energy is required but in order to obtain this one
should have a definite idea of the analytic structure of 
the distribution of eigenvalues.

The calculations we have performed here are  not only relevant to string 
theory in one dimension but also for other problems related 
to the matrix chain.
An example is  
the Kazakov-Migdal lattice model of induced QCD in $D$-dimensions \cite{kmm}.
Devised as a lattice model of gauge interactions where the self-interaction 
of adjoint scalar fields $\phi_x$ are supposed to induce the standard Wilson 
term which is left out, it is defined by the following action
\beq
S_{KM}= \sum_x \Tr V(\phi_x) - \sum_{\la xy \ra} \Tr \Lambda_{xy} \phi_y 
\Lambda_{yx} \phi_x
\label{kmmaction}
\eeq
Here the sub-scripts label sites on a hyper-cubic lattice on which 
the $N \times N$ adjoint scalar fields $\phi_x$ reside.   Nearest
neighbour interactions are mediated by $SU(N)$ gauge fields $\Lambda$ residing
on the lattice links.
From the large $N$ saddle-point equations it can be shown \cite{mat1}
that
the one-dimensional matrix chain we have considered (\ref{mcaction}), 
with potential $U$,  is equivalent to the Kazakov-Migdal model with 
a potential satisfying
\beq
V^\prime(x) =  D U^\prime(x) - 2 (D-1) \pvint dz \frac{\rho_0(z)}{x-z}
\eeq
where the eigenvalue density $\rho_0$ is common to both models.

The one drawback of this model of induced QCD is that the action
(\ref{kmmaction}) has a $\IZ_N$
symmetry under the center of the gauge group
which leads to super-confinement of gauge degrees of freedom, 
even on the lattice scale \cite{ksw}.  Obviously this not 
a feature that one would like to have in the continuum limit
of the model and the easiest way to avoid it is for the system
not to realize this symmetry faithfully in its solution.
Unfortunately for the potentials $U$ of logarithmic form, 
for which we have $\rho_0$ explicitly (see Section Four),
this symmetry persists \cite{penner,dmsw}. 
It would be interesting to see if solutions of the matrix chain 
for more general potentials might lead to 
non-trivial propagation of gauge fields in the 
continuum limit of the Kazakov-Migdal model.
This can be checked in principle using the solution of the matrix chain for 
the eigenvalue density $\rho_0$ and the formalisms of \cite{shatash} and 
\cite{morozov} which express the correlation of gauge fields in the 
model in terms of $\rho_0$.

\section*{Acknowledgments}
We would like to thank S. Jaimungal, V. Kazakov, I. Klebanov and
E. Lieb
for helpful comments and the Niels Bohr Institute for its 
hospitality during a visit in which this work was begun.  
This work was completed with the  
support of the Natural Sciences and Engineering Research Council of Canada.

\end{document}